\RequirePackage{fix-cm}
\documentclass[twocolumn]{svjour3}          
\smartqed  
\usepackage{graphicx}
\usepackage{indentfirst}
\usepackage{amsmath}
\usepackage{amsfonts}
\usepackage{array}
\usepackage{multirow}
\usepackage{tabularx}
\usepackage{comment}
\begin{document}

\title{Synchronization of coherence resonance oscillators subject to L{\'e}vy noise%
}


\author{lvan~A.~Korneev         \and
        Vladimir V. Semenov 
}


\institute{
I.A. Korneev  \at 
Saratov State University, Astrakhanskaya str. 83, 410012, Saratov, Russia \\
           	\and
V.V. Semenov \at 
Saratov State University, Astrakhanskaya str. 83, 410012, Saratov, Russia \\
              \email{semenov.v.v.ssu@gmail.com}    
}

\date{Received: date / Accepted: date}

\maketitle

\begin{abstract}
Using methods of numerical simulation, we analyze the influence of L{\'e}vy noise on synchronization of excitable oscillators in the regime of coherence resonance. Three cases are under consideration: forced synchronization of a single FitzHugh-Nagumo oscillator subject to periodic forcing, mutual synchronization of two coupled FitzHugh-Nagumo oscillators and ensembles of locally and globally coupled FitzHugh-Nagumo neurons. It is demonstrated that L{\'e}vy noise provides for transformation of the forced synchronization area such that synchronization can arise or be destroyed through varying the L{\'e}vy noise parameters at fixed frequency and amplitude of the external force. Moreover, the L{\'e}vy noise intrinsic peculiarity can induce the counterintuitive transformation of the synchronization areas such that increasing the external force amplitude gives rise to leaving the synchronization area. In the context of synchronization of coupled oscillators, L{\'e}vy noise is also shown to control the transition to synchronization which can be achieved at lower or higher values of the coupling strength when changing the L{\'e}vy noise parameters. However, such effects are found to be exhibited by ensembles of coupled oscillators, whereas the influence of L{\'e}vy noise on mutual synchronization of two coupled coherence resonance oscillators is minimal and does not lead to significant changes as compared to Gaussian noise. 
\end{abstract}

\section{Introduction}
\label{intro}
The essence of coherence resonance consists in increasing the noise-induced oscillation regularity for an optimum value of the noise intensity which can be achieved in a wide spectrum of excitable \cite{pikovsky1997,lindner1999,lindner2004,deville2005,muratov2005,semenov2017,semenov2018} and non-excitable \cite{gang1993,ushakov2005,zakharova2010,zakharova2013,geffert2014,semenov2015} systems. The manifestations of coherence resonance are reported to be observed when analyzing stochastic processes in neurodynamics \cite{pikovsky1997,lee1998,lindner2004,pisarchik2023,tateno2004}, microwave \cite{dmitriev2011} and semiconductor \cite{hizanidis2006,huang2014,shao2018} electronics, optics \cite{dubbeldam1999,giacomelli2000,avila2004,otto2014,arteaga2007,arecchi2009}, quantum physics \cite{kato2021}, thermoacoustics \cite{kabiraj2015}, plasma physics \cite{shaw2015}, hydrodynamics \cite{zhu2019}, climatology \cite{bosio2023} and chemistry \cite{miyakawa2002,beato2008,simakov2013}. The occurrence of coherence resonance can be accompanied by another effects, for instance, synchronization \cite{balanov2009}. In such a case, the phenomenon of synchronization is associated with the similarity of the noise-induced dynamics of excitable systems in the regime of coherence resonance with the self-oscillatory behaviour. It is known that noise-induced oscillations corresponding to coherence resonance can be synchronized mutually or by external forcing \cite{han1999,ciszak2003,ciszak2004}. Moreover, the synchronization of the noise-induced oscillations occurs in a similar way as for a deterministic quasiperiodic system \cite{astakhov2011}.

A broad variety of dynamical systems, processes and associated applications dictates the significance of coherence resonance control in the context of both fundamental and applied science. Various methods can be applied for this purpose. For instance, introduction of time-delayed feedback allows to control the characteristics of noise-induced oscillations in systems with type-I \cite{aust2010} and type-II \cite{janson2004,brandstetter2010} excitability as well as in non-excitable systems \cite{geffert2014,semenov2015} exhibiting the effect of coherence resonance. In networks of coupled oscillators, one can use the coupling properties for controlling coherence resonance. In particular, one can vary the coupling strength and modify the coupling topology which was successfully demonstrated on examples of multilayer networks with multiplexing \cite{semenova2018,masoliver2021}. The third approach consists in adjusting the properties of noise to enhance or to suppress coherence resonance. Such effects can be realized when varying the correlation time of coloured noise \cite{brandstetter2010}. As demonstrated both numerically and experimentally in recent paper \cite{semenov2024}, coherence resonance in excitable systems can be efficiently controlled by varying the L{\'e}vy noise parameters. Based on this result, in the present paper we study the effect of L{\'e}vy noise on synchronization of excitable oscillators in the regime of coherence resonance. Generally, the idea testified in current research consists in the assumption that it is more or less difficult to synchronize the suppressed (less regular) or enhanced (more regular) coherence resonance dynamics induced at different sets of the L{\'e}vy noise parameters.

Characterized by the appearance of large, potentially infinite, jumps, L{\'e}vy noise\footnote{The term 'L{\'e}vy noise' is used to distinguish a class of stable non-Gaussian noise that exhibits long heavy tails of its distribution of the probability density function.} is an appropriate model to describe the stochastic dynamics associated with abrupt changes are related effects, which can be observed in lasers \cite{rocha2020}, cardiac rhythms \cite{peng1993}, molecular motors \cite{lisowski2015}, quantum dots \cite{novikov2005}, financial \cite{mantegna1999,barndorff2001} and social \cite{perc2007} systems. In the context of biological neural networks, stochastic processes with a L{\'e}vy distribution are often more accurate model as compared to Gaussian noise \cite{nurzaman2011,wu2017}. 

In the context of noise-induced phenomena, L{\'e}vy noise is known to provide for increasing in neuronal firings  \cite{vinaya2018}, disappearance of the winner-take-all behaviour in neural competition models \cite{feng2019}, noise-sustained detection of faint or subthreshold signals \cite{patel2008}. Moreover, L{\'e}vy noise enables controlling characteristics of noise-induced oscillations in the regime of conventional \cite{dybiec2006,dybiec2009,yonkeu2020} and self-induced \cite{yamakou2022} stochastic resonance, coherence resonance in excitable \cite{semenov2024} and non-excitable systems \cite{yonkeu2020}, stochastic wavefront propagation associated with the property of bistability \cite{semenov2025} and excitability \cite{korneev2024}. The current paper extends this list by noise-induced synchronization of spiking dynamics and by the destruction of the stochastic synchronization due to the intrinsic peculiarities of L{\'e}vy noise.

\section{Single oscillator subject to periodic forcing}
We begin our research starting from a single non-autonomous FitzHugh-Nagumo oscillator \cite{fitzhugh1961,nagumo1962} being a paradigmatic model for the type-II excitability and considered in the simplest form:
\begin{equation}
\label{eq:single_system} 
\begin{array}{l} 
\varepsilon\dfrac{dx}{dt} = x-x^3/3-y,\\ 
\dfrac{dy}{dt}=x+a+A_{\text{ext}}\sin(\omega_{\text{ext}} t)+\xi(t), 
\end{array}
\end{equation} 
where $x=x(t)$ and $y=y(t)$ are dynamical variables. A parameter $\varepsilon\ll 1$ is responsible for the time scale separation of fast activator, $x$, and slow inhibitor, $y$, variables, $a$ is the threshold parameter which determines the system dynamics: the system exhibits the excitable regime at $|a|>1$ and the oscillatory one for $|a|<1$. In this paper, single and coupled FitzHugh-Nagumo neurons are considered in the excitable regime ($a=1.05$ and $\varepsilon=0.05$) for varying parameters of additive L{\'e}vy noise $\xi(t)$ (defined as the formal derivative of the L{\'e}vy stable motion) and external periodic forcing (amplitude $A_{\text{ext}}$ and  frequency $\omega_{\text{ext}}$). 

L{\'e}vy noise is characterized by four parameters: stability index $\alpha \in (0:2]$ (case $\alpha=1$ is excluded from the consideration in the current research), skewness (asymmetry) parameter $\beta\in [-1:1]$, mean value $\mu=0$ ($\mu$ is set to be zero for the strictly stable distributions \cite{janicki1994}) and scale parameter $\sigma$. Parameter $D=\sigma^{\alpha}$ is introduced as the noise intensity. If $\xi(t)$ obeys to L{\'e}vy distribution $L_{\alpha,\beta}(\xi,\sigma,\mu)$, its characteristic function takes the form \cite{janicki1994,dybiec2006,dybiec2007}:
\begin{equation}
\label{eq:characteristic_function}
\begin{array}{l} 
\phi(k)=\int\limits_{-\infty}^{+\infty}\exp(ikx)L_{\alpha,\beta}(\xi,\sigma,\mu)dx,\\
=\exp\left[-\sigma^{\alpha}|k|^{\alpha}\left(1-i\beta sgn(k)\tan\dfrac{\pi\alpha}{2} \right) \right].
\end{array}
\end{equation}
To generate random sequence $\xi$ corresponding to characteristic function (\ref{eq:characteristic_function}), the Janicki-Weron algorithm is used \cite{janicki1994,weron1995}: 
\begin{equation}
\label{eq:noise_generation} 
\begin{array}{l} 
\xi=\sigma S_{\alpha,\beta}\times \dfrac{\sin(\alpha(V+B_{\alpha,\beta}))}{(\cos(V))^{1/\alpha}}\\
\times \left( \dfrac{\cos(V-\alpha(V+B_{\alpha,\beta}))}{W}\right)^{\dfrac{1-\alpha}{\alpha}},
\end{array}
\end{equation} 
where constants are $B_{\alpha,\beta}=\left( \text{arctan} \left( \beta \tan \left( \dfrac{\pi\alpha}{2}\right)\right) \right)/\alpha$ and $S_{\alpha,\beta}=\left( 1+\beta^2 \tan ^2\left( \dfrac{\pi\alpha}{2}\right)\right)^{1/2\alpha}$, $V$ is a random variable uniformly distributed in the range $\left(-\dfrac{\pi}{2}:\dfrac{\pi}{2}\right)$, $W$ is an exponential random variable with mean 1 (variables $W$ and $V$ are independent). In case $\alpha=2$, the distribution of the probability density function (PDF) takes the Gaussian form with zero mean value and the variance being equal to $2\sigma^2$. If $\alpha<2$, the distribution is non-Gaussian and the variance becomes infinite.

Numerical simulations of single oscillator (\ref{eq:single_system}) and coupled oscillators below are carried out by integration using the Heun method \cite{mannella2002} with the time step $\Delta t=10^{-3}$ or smaller. It is important to note that numerical modelling of equations including $\alpha$-stable stochastic process with finite time step implies the normalization of the noise term by $\Delta t^{1/\alpha}$ \cite{xu2016,pavlyukevich2010}.

\begin{figure}
\centering
\includegraphics[width=0.48\textwidth]{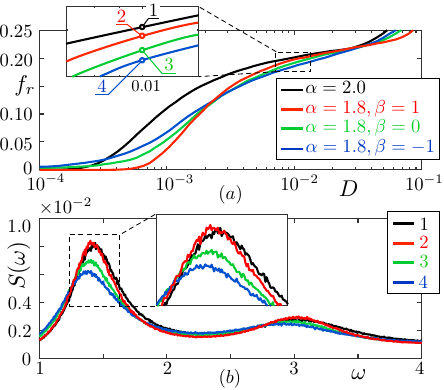} 
\caption{Evolution of the L{\'e}vy-noise-induced spiking activity when varying the noise parameters in single FitzHugh-Nagumo oscillator (\ref{eq:single_system}) in the absence of external forcing ($A_{\text{ext}}\equiv 0$): the dependencies of the firing rate (see Eq. (\ref{eq:single_firing_rate})) on the noise intensity for varying $\alpha$ and $\beta$ (panel (a)) and the power spectra of oscillations $x(t)$ in control points 1-4 corresponding to $D=0.01$ (panel (b)). The oscillator's parameters are $a=1.05$, $\varepsilon=0.05$.}
\label{fig1}
\end{figure}  

In the absence of external periodic forcing (at $A_{\text{ext}}\equiv 0$), the impact of L{\'e}vy noise results in the occurrence of coherence resonance with controllable characteristics when varying the L{\'e}vy noise parameters $\alpha$ and $\beta$ (see Ref. \cite{semenov2024}). In addition to control of the noise-induced oscillation regularity, such approach allows to affect the neuronal firing activity. To illustrate this fact, the dependence of the spike firing frequency on the noise intensity is used, where the firing rate is introduced to be proportional to a number of spikes $M$ induced by noise during the time period $t\in[0:T]$:
\begin{equation}
\label{eq:single_firing_rate} 
f_r=\lim_{T\to\infty}\dfrac{M(T)}{T}.
\end{equation} 
As depicted in Fig.~\ref{fig1}~(a), one can change the firing rate by tuning the noise parameters $\alpha$ and $\beta$ at fixed noise intensity. This is also reflected in the evolution of the power spectrum where the main spectral peak shifts slightly for varying $\alpha$ and $\beta$ and fixed $D=0.01$. 

In the context of the power spectrum evolution, the intrinsic peculiarities of the L{\'e}vy noise can provide for enhancing and suppressing coherence resonance (this result correlates with materials of Ref. \cite{semenov2024}) manifested as transformations of the main spectral peak such that it becomes more and less pronounced (i.e., the ratio of the height of the main peak in the power spectrum to its width changes) as well as for shift of the main spectral peak frequency. The important detail must be emphasized: the rescaled spike firing frequency $2\pi f_r$ in points 1-4 in Fig.~\ref{fig1}~(a) (varies in the range $f_r\in[1.16:1.26]$) is not equal to the frequency of the main spectral peak ($\omega_{\text{peak}} \in [1.36:1.45]$). This is due to the fact that the noise-induced spiking dynamics represents complex anharmonic process. 

In the presence of external harmonic force, one can observe synchronization of noise-induced oscillations in the regime of coherence resonance. To quantitatively describe this effect, the frequency ratio is introduced as
\begin{equation}
\label{eq:single_frequency_ratio} 
R_f=\dfrac{2\pi f_r}{\omega_{\text{ext}}}-1.
\end{equation} 
Then quantity $R_f$ is analyzed as a function of $\omega_{\text{ext}}$ and $A_{\text{ext}}$. In addition, we take into consideration the realizations of phases, where the instantaneous oscillation phase is determined from the results of numerical simulation of model (\ref{eq:single_system}) as $\varphi(t)=\text{arctan}\left( y(t)/x(t)\right)\pm \pi k$ ($k$ is an integer variable determined by the requirement of the phase's continuity). Then the average difference frequency $\Omega$ is introduced as an additional quantity to characterise the stochastic synchronization:
\begin{equation}
\label{eq:single_beat_frequency} 
\Omega=\lim_{T\to\infty}\dfrac{\varphi(T)-\omega_{\text{ext}}T}{T}.
\end{equation} 

\begin{figure*}
\centering
\includegraphics[width=0.75\textwidth]{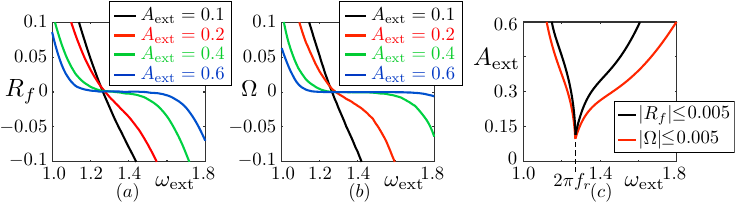} 
\caption{Forced synchronization of stochastic oscillations in coherence resonance oscillator (\ref{eq:single_system}) subject to Gaussian noise ($\alpha=2$) manifested as the evolution of the dependencies  $R_f(\omega_{\text{ext}})$ (panel(a)) and $\Omega(\omega_{\text{ext}})$ (panel (b)) when the  amplitude of harmonic forcing, $A_{\text{ext}}$, growths. Panel (c) depicts the areas of obeying the condition $R_f \leq 0.005$ (black boundaries) and $\Omega \leq 0.005$ (red boundaries) in parameter plane ($\omega_{\text{ext}}$,$A_{\text{ext}}$). The oscillator's parameters are $a=1.05$, $\varepsilon=0.05$, the noise intensity is $D=0.01$.}
\label{fig2}
\end{figure*}  

First, the effect of synchronization is studied in the presence of Gaussian noise (Eqs. (\ref{eq:single_system}) at $\alpha=2$) of fixed intensity $D=0.01$. In the presence of weak external forcing, i.e. small $A_{\text{ext}}$, the transition to the synchronous state is not indicated: there is no finite range of $\omega_{\text{ext}}$ where $R$ and $\Omega$ tend to zero (see the curves $R_f(\omega_{\text{ext}})$ and $\Omega(\omega_{\text{ext}})$ corresponding $A_{\text{ext}}=0.1$ in Fig.~\ref{fig2}~(a),(b)). When increasing amplitude $A_{\text{ext}}$, both curves transform such that one can distinguish a finite range where  quantities $R_f$ and $\Omega$ approach zero. Thus, at sufficiently high amplitude $A_{\text{ext}}$ one can easily identify within a certain range of the external forcing frequency $\omega_{\text{ext}}$, where $R_f$ and $\Omega$ possess extremely small values, i.e. the frequency entrainment effect is realized. The frequency-locked interval tends to become broader as the external forcing amplitude is increased such that there are the triangular-shaped zones which resemble the well-known Arnold tongues in ($\omega_{\text{ext}}$,$A_{\text{ext}}$) parameter plane where the frequencies of noise-induced oscillations are locked [Fig. \ref{fig2}~(c)]. More precisely, the expressions $\Omega \leq 0.005$ and $R_f \leq 0.005$ were used to diagnose whether the frequencies were locked (value $0.005$ is considered as a threshold level). 

It is important to note the subjective character of choosing the threshold values of $\Omega$ and $R_f$ used for estimation of the synchronization area boundaries. Indeed, the boundaries transform when changing the threshold values for $\Omega$ and $R_f$. Moreover, the boundaries of the synchronization area obtained by means of analysis of $\Omega$ and $R_f$ differ from each other. In addition, estimation of the synchronization by means of the spectral analysis is also subjective and will result in the third option for the boundaries of the synchronization area. In such a case, the difference between results of applying various approaches becomes especially pronounced in the neighbourhood of the normalized firing frequencies $2\pi f_r$ and the frequency $\omega_{\text{peak}}$ corresponding to the main spectral peak of the autonomous oscillator (as noted above, these frequencies differ from each other). In the current research, we are focused on the exploration of synchronization in the context of the quantities $\Omega$ and $R_f$, whereas study of the forced synchronization in the context of spectral analysis is a subject for further study. 

\begin{figure}
\centering
\includegraphics[width=0.48\textwidth]{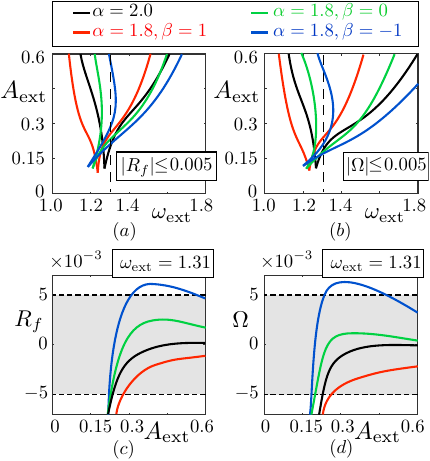} 
\caption{Control of forced synchronization in single FitzHugh-Nagumo oscillator (\ref{eq:single_system}) by varying the L{\'e}vy noise parameters: Panels (a) and (b) depict boundaries of the synchronization tongues estimated by means of conditions $\left| R_f \right| \leq 0.005$ (panel (a)) and  $\left| \Omega \right|\leq 0.005$ (panel (b)). Panels (c) and (d) demonstrate the dependencies $R_f (A_{\text{ext}})$ (panel (c)) and $\Omega (A_{\text{ext}})$ (panel (d)) at fixed frequency $\omega_{\text{ext}}=1.31$ (correspond to the dashed lines in panels (a) and (b)). The oscillator's parameters and the noise intensity are the same as in the previous figure. All the curves are obtained for four sets of the L{\'e}vy noise parameters: $\alpha=2$ (black curves), $\alpha=1.8$ and $\beta=1$ (red curves), $\alpha=1.8$ and $\beta=0$ (green curves), $\alpha=1.8$ and $\beta=-1$ (blue curves).}
\label{fig3}
\end{figure}

To characterise the evolution of the stochastic dynamics caused by varying the L{\'e}vy noise parameters, we use the same conditions $\Omega \leq 0.005$ and $R_f \leq 0.005$ as compared to the presence of Gaussian noise. As demonstrated in Ref. \cite{semenov2024}, coherence resonance in model (\ref{eq:single_system}) is enhanced at $\alpha=1.8$, $\beta=1$ and suppressed at $\alpha=1.8$, $\beta=-1$ and $\alpha=1.8$, $\beta=0$. In the following, the same sets of parameters are used to compare the aspect of synchronization in the presence of Gaussian ($\alpha=2$) and L{\'e}vy noise. As depicted in Fig.~\ref{fig3} (a),(b) one can shift the boundaries of the synchronization area when varying noise parameters $\alpha$ and $\beta$, which is exhibited by both quantities $\Omega$ and $R_f$. In particular, in case $\alpha<1.8$ one can shift the synchronization area to the left and to the right by tuning parameter $\beta$. At the same time, the synchronization tongue base (the boundaries of the synchronization area at weak external forcing amplitudes) shifts to the new frequency value $2\pi f_r$ corresponding to the firing rate achieved in the autonomous oscillator at chosen parameters of L{\'e}vy noise. 

The intrinsic peculiarity of the stochastic synchronization in the presence of L{\'e}vy noise consists in the warped shape of the synchronization area such that one can get in and out the areas when increasing the periodic force amplitude at fixed frequency (see the dashed lines in Fig.~\ref{fig3}~(a),(b)). A paradoxical effect occurs and becomes more pronounced at lower values of $\beta$: increasing the amplitude of the external harmonic influence can make the oscillations less synchronized. Considering the curves  $R_f (A_{\text{ext}})$ [Fig.~\ref{fig3}~(c)] and $\Omega (A_{\text{ext}})$ [Fig.~\ref{fig3}~(d)] at certain fixed values of frequency $\omega_{\text{ext}}$ allows to reveal 
the key property of the forced synchronization of the L{\'e}vy-noise-induced dynamics in the regime of coherence resonance: the phase and frequency differences between external forcing and driven system finally tend to zero when increasing the external forcing amplitude, but can behave non-monotonically and increase during this process.

\section{Two coupled oscillators}
\begin{figure}
\centering
\includegraphics[width=0.48\textwidth]{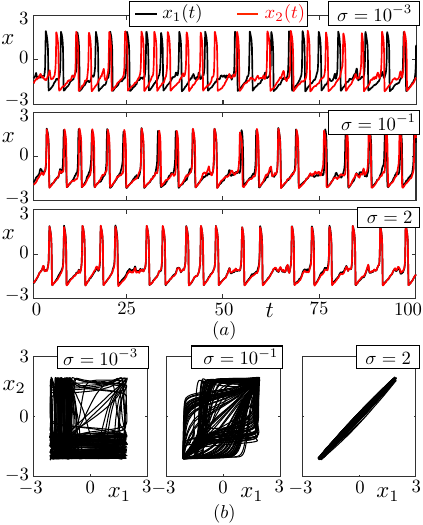} 
\caption{Transition to mutual synchronization in coupled coherence resonance oscillators (\ref{eq:two_systems}) subject to Gaussian noise ($\alpha=2$) when increasing the coupling strength reflected as the evolution of time realizations $x_{1,2}(t)$ (panel (a)) and phase portraits in ($x_1$,$x_2$) phase plane. The oscillators' parameters are $a=1.05$, $\varepsilon=0.05$, the noise intensity is $D=0.01$.}
\label{fig4}
\end{figure}

Let us consider the effect of mutual synchronization on an example of two coupled stochastic excitable oscillators:
\begin{equation}
\label{eq:two_systems} 
\begin{array}{l} 
\varepsilon\dfrac{dx_1}{dt} = x_1-x_1^3/3-y_1+\sigma(x_2-x_1),\\ 
\dfrac{dy_1}{dt}=x_1+a+\xi_1(t),\\
\varepsilon\dfrac{dx_2}{dt} = x_2-x_2^3/3-y_2+\sigma(x_1-x_2),\\ 
\dfrac{dy_2}{dt}=x_2+a+\xi_2(t),\\ 
\end{array}
\end{equation} 
where $\sigma$ is the coupling strength, $\xi_{1,2}(t)$ are statistically independent sources of L{\'e}vy noise generated according to algorithm (\ref{eq:noise_generation}), the oscillators' parameters are the same as in the previous section, $\varepsilon=0.05$, $a=1.05$. Oscillators (\ref{eq:two_systems}) seem to be identical, but spontaneously produce spikes in different time moments due to the action of the noise sources. In the presence of weak coupling, such process occurs independently which is clearly visible in time realizations $x_{1,2}(t)$ [Fig.~\ref{fig4}~(a), upper panel] and phase portraits in ($x_1$,$x_2$) phase plane [Fig.~\ref{fig4}~(b), left panel]. Increasing the coupling strength continuously makes the spiking activity more and more synchronous (compare the panels in Fig.\ref{fig4} corresponding to $\sigma=10^{-3}$ and $\sigma=10^{-1}$). Finally, growth of the coupling strength leads to the transformation such that the oscillations of two oscillators almost coincide in time (see the lower panel in Fig.~\ref{fig4}~(a) and right panel in Fig.~\ref{fig4}~(b)) and the dynamics tends to the regime of complete synchronization. 

\begin{figure}
\centering
\includegraphics[width=0.48\textwidth]{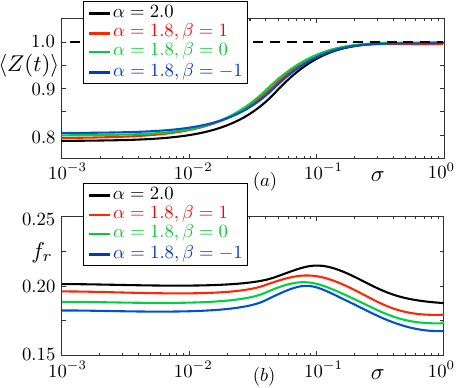} 
\caption{Dependencies of the mean order parameter (panel (a)) and the firing rate (panel (b)) on the coupling strength obtained in model (\ref{eq:two_systems}) at different values of the L{\'e}vy noise parameters. The oscillators' parameters and the noise intensity are the same as in the previous figure.}
\label{fig5}
\end{figure}  

To quantitatively describe the transition to the synchronous oscillations and the impact of L{\'e}vy noise on this process, we introduce the mean value of the order parameter: 
\begin{equation}
\label{eq:pair_order_parameter} 
\left< Z(t) \right>=\left< \dfrac{1}{2} \left| \exp (j\varphi_1(t))+\exp(j\varphi_2(t)) \right| \right>,
\end{equation} 
where the geometric phases $\varphi_{1,2}(t)$ of the interacting oscillators are defined in the same way as compared to the single oscillator (see the previous section), $j$ is the imaginary unit, brackets $\left< ... \right>$ mean the value averaged over time. Varying in the range $[0:1]$, the quantity $\left< Z(t) \right>$ increases when the transition to synchronization occurs (values $\left< Z(t) \right>\approx 1$ and $\left< Z(t) \right> <1$ indicate synchronous and asynchronous oscillations, respectively). In the following, $\left< Z(t) \right>$ is considered as a function of the coupling strength at different parameters of L{\'e}vy noise. In addition, the spike firing rate $f_r$ is analysed in the similar way. The corresponding results are depicted in Fig. \ref{fig5} and clearly indicate that L{\'e}vy noise has no principal impact on the mutual synchronization of noise-induced oscillations. Indeed, the curves $\left< Z(t) \right>$ obtained for different values of the L{\'e}vy noise parameters [Fig.~\ref{fig5}~(a)] are not much different from each other as well as the dependencies of the firing rate on the coupling strength [Fig.~\ref{fig5}~(b)]. However, the situation changes when the number of interacting oscillators increases.

\section{Ensembles of coupled oscillators}
The third system under study represents an ensemble of coupled FitzHugh-Nagumo oscillators in the excitable regime:
\begin{equation}
\label{eq:ensemble} 
\begin{array}{l} 
\varepsilon\dfrac{dx_i}{dt} = x_i-x_i^3/3-y_i+g_i(x_1,x_2,...,x_N),\\ 
\dfrac{dy_i}{dt}=x_i+a+\xi_i(t),
\end{array}
\end{equation} 
where $i=1,2,3,...,N$ ($N$ is the total number of oscillators), $\varepsilon=0.05$, $a=1.05$, $\xi_i(t)$ are statistically independent sources of the L{\'e}vy noise generated according to algorithm (\ref{eq:noise_generation}), $g_i(x_1,x_2,...,x_N)$ is the coupling term. The consideration starts from the case of global coupling:
 \begin{equation}
\label{eq:global_coupling}
\begin{array}{l}
g_i(x_1,x_2,...,x_N)= \dfrac{\sigma}{N} \sum\limits_{j=1}^{N}(x_j-x_i),
\end{array}
\end{equation}
where $\sigma$ is the coupling strength. Numerically obtained time realizations $x_i(t)$ are used to build spatio-temporal diagrams visualising the collective dynamics. In addition, time realizations $x_i(t)$ and $y_i(t)$ are involved into the calculations of the global order parameter at all the time moments $t$ for further extraction of the mean value. The mean value of the global order parameter is introduced in the similar form as compared to the previous section where a pair of interacting oscillators is discussed (see Eq. (\ref{eq:pair_order_parameter})):
\begin{equation}
\label{eq:global_order_parameter_ensemble}
\begin{array}{l}
\left< Z(t) \right>=\left< \dfrac{1}{N} \left| \sum\limits_{i=1}^{N}\exp (j\varphi_i(t)) \right| \right>,
\end{array}
\end{equation}
where $\varphi_i(t)=\text{arctan}\left( y_i(t)/x_i(t)\right)\pm \pi k_i$. Similarly to the study of single oscillator (\ref{eq:single_system}) and two coupled oscillators (\ref{eq:two_systems}) presented above, $k$ is an integer variable introduced to obey the requirement of the phase's continuity.

\begin{figure}
\centering
\includegraphics[width=0.5\textwidth]{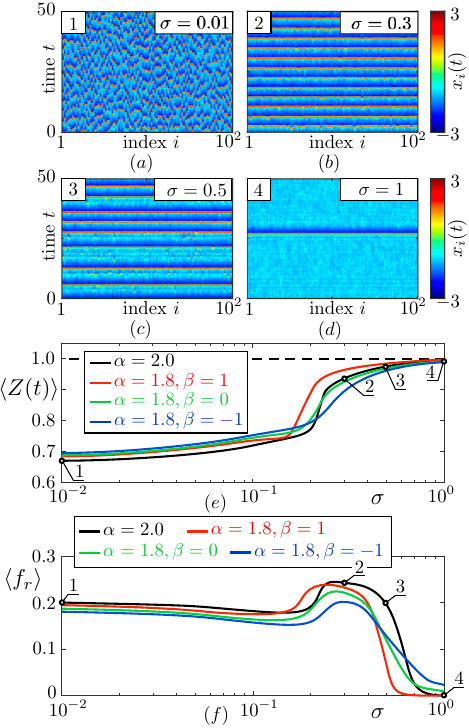} 
\caption{Stochastic synchronization in ensemble (\ref{eq:ensemble}) in the presence of global coupling (\ref{eq:global_coupling}): (a)-(d) Space-time plots illustrating synchronization and further suppression of Gaussian-noise-induced ($\alpha=2$) spiking activity when increasing the coupling strength; (e)-(f) Dependencies of the mean global order parameter (panel (e)) and the averaged firing rate (panel (f)) on the coupling strength at different values of the L{\'e}vy noise parameters. Points 1-4 in panels (e) and (f) correspond to space-time plots in panels (a)-(d). The oscillators' parameters are $a=1.05$, $\varepsilon=0.05$, the noise intensity is $D=0.01$.}
\label{fig6}
\end{figure}  

In the presence of weak coupling, the collective stochastic dynamics in ensemble (\ref{eq:ensemble})-(\ref{eq:global_coupling}) is asynchronous, the spiking activity of interacting oscillators has spontaneous, independent character which is clearly visible in the spatio-temporal diagrams [Fig.~\ref{fig6}~(a)]. When increasing the coupling strength, the transition to synchronization occurs such that the oscillations are in phase [Fig.~\ref{fig6}~(b)]. After the regime of synchronization is achieved, further growth of the coupling strength suppresses the spiking activity such that the spikes become less and less frequent [Fig.~\ref{fig6}~(c),(d)]. Quantitatively, the described evolution is reflected in the dependencies of the global order parameter [Fig.~\ref{fig6}~(e)] and the firing rate averaged over the ensemble [Fig.~\ref{fig6}~(f)] on the coupling strength obtained for different values of the L{\'e}vy noise parameters. 

Two aspects distinguish the synchronization of two coupled oscillators (\ref{eq:two_systems}) and synchronization in ensemble of globally coupled oscillators (model (\ref{eq:ensemble}) for the coupling strength (\ref{eq:global_coupling})). Firstly, L{\'e}vy noise becomes a factor for controlling the synchronization if the number of interacting oscillators is large enough. In particular, varying parameters $\alpha$ and $\beta$, one can realize transition to the synchronous dynamics in ensemble (\ref{eq:ensemble}) at lower or larger values of the coupling strength (compare Fig.~\ref{fig5}~(a) and Fig.~\ref{fig6}~(e)). The second revealed feature of the collective dynamics in the ensemble consists in decreasing the averaged firing rate up to extremely low values, which was not observed in a pair of coupled oscillators (\ref{eq:two_systems}) (compare Fig.~\ref{fig5}~(b) and Fig.~\ref{fig6}~(f))). 

\begin{figure}
\centering
\includegraphics[width=0.5\textwidth]{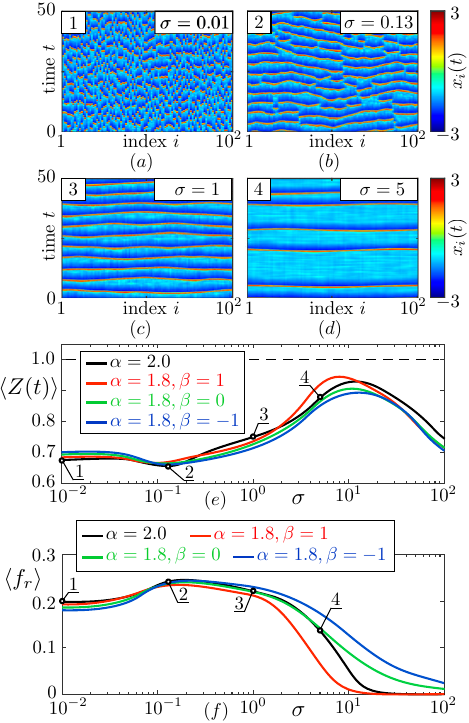} 
\caption{Stochastic synchronization in ensemble (\ref{eq:ensemble}) in the presence of local coupling (\ref{eq:local_coupling}): (a)-(d) Space-time plots illustrating synchronization and further suppression of Gaussian-noise-induced ($\alpha=2$) spiking activity when increasing the coupling strength; (e)-(f) Dependencies of the mean global order parameter (panel (e)) and the averaged firing rate (panel (f)) on the coupling strength at different values of the L{\'e}vy noise parameters. Points 1-4 in panels (e) and (f) correspond to space-time plots in panels (a)-(d). The oscillators' parameters and the noise intensity are the same as in the previous figure.}
\label{fig7}
\end{figure}

In the presence of local coupling, i.e. when the coupling term is described as 
\begin{equation}
\label{eq:local_coupling}
\begin{array}{l}
g_i(x_1,x_2,...,x_N)= \dfrac{\sigma}{2} (x_{i-1}+x_{i+1}-2x_i),
\end{array}
\end{equation}
the transition to synchronization results in formation of spatially coherent dynamics such that the spiking activity of neighbour oscillators is almost in phase (see the space-time plots in Fig.~\ref{fig7}~(a)-(d)), but the fluctuations of phases persist in general and the corresponding space time-plot are less aligned as compared to the ones obtained for the global coupling (compare Fig.~\ref{fig6}~(a)-(d) and Fig.~\ref{fig7}~(a)-(d)). In such a case, one achieves lower values of the global order parameter when realizing the coherent oscillatory dynamics (maximal values of $\left< Z(t)\right>$ are around 0.94).  Moreover, the global order parameter can behave non-monotonically in the presence of local coupling such that one can observe decreasing the global order parameter with growth of the coupling strength [Fig.~\ref{fig7}~(e)]. The impact of local [Fig.~\ref{fig7}~(f)] and global [Fig.~\ref{fig6}~(f)] coupling on the firing rate has the same character: if the coupling strength is large enough, the spiking activity is suppressed which leads to the exhibition of the  quiescent regime characterized by fluctuations in the neighbourhood of the steady state.

In the context of the influence on synchronization, L{\'e}vy noise similarly acts on the collective dynamics in ensemble (\ref{eq:ensemble}) in the presence of coupling terms (\ref{eq:global_coupling}) and (\ref{eq:local_coupling}). The same sets of the L{\'e}vy noise parameters similarly bring the moment of synchronization closer or distance the moment of synchronization when the coupling strength increases in the ensemble with local or global coupling. 

\section*{Conclusions}
\label{conclusions}
As demonstrated in the current paper on an example of coherence resonance oscillators, L{\'e}vy noise represents a promising tool for controlling the effect of synchronization. 
 In particular, varying the L{\'e}vy noise parameters allows to transform the areas of forced synchronization. This indicates that one can induce or destroy forced synchronization of a single spiking oscillator at fixed, changeless frequency and amplitude of the external harmonic force. Moreover, the sinchronization areas can be reshaped such that it becomes possible to leave the synchronization area by increasing the external force amplitude at fixed frequency of the external impact. To be sure that this intriguing aspect takes place and does not result from inaccuracies of the numerical simulation, two independent quantities, the firing rate and the average difference frequency were introduced to identify the occurrence of synchronization.

In the context of coupled oscillators, the impact of L{\'e}vy noise on synchronization depends on the number of interacting elements. In more detail, varying the L{\'e}vy noise parameters does not lead to significant changes in the dynamics of two coupled coherence resonance oscillators. In contrast, L{\'e}vy noise becomes a control factor when one deals with ensembles of coupled oscillators. In such a case, synchronization arise earlier or later, which depends on the peculiarities of the stochastic impact. The comparative analysis of the results obtained in the present paper and materials of Ref. \cite{semenov2024} allows to notice that the sets of the L{\'e}vy noise parameters corresponding to enhancement of coherence resonance also provide for achieving the synchronization of noise-induced spiking at lower coupling strength. Similarly, if the coherence resonance in an excitable oscillator is less pronounced at chosen parameters of L{\'e}vy noise, then synchrotinaztion of such oscillators occurs at higher values of the coupling strength. In other words, the more regular is the individual dynamics of the excitable oscillators in an ensemble, the less intensive interaction gives rise to synchronous collective dynamics. To emphasize general character of the L{\'e}vy noise impact on the collective behaviour of spiking oscillators, the L{\'e}vy-noise-based control of synchronization was demonstrated for two kinds of the coupling topology: local and global coupling.

A distinguishable effect was revealed in ensembles of globally and locally coupled excitable oscillators. If the intensity of interaction in ensembles becomes large enough, further increasing the coupling strength suppress the spiking activity such that the spikes are less and less frequent up to extremely low values of the firing rate. Certain ranges of the coupling strength values where such effects take place depend on the coupling topology and parameters of L{\'e}vy noise.

The presented results are the basis for further studies such as experimental considerations of the observed phenomena by means of electronic modelling \cite{semenov2024_book} similarly to papers \cite{semenov2024,semenov2025}. In addition, it is not clear how the intrinsic peculiarities of L{\'e}vy noise impact  coherence resonance and synchronization in a case of more complicated dynamics. In particular, the influence of L{\'e}vy noise on coherence resonance oscillators exhibiting the neuronal bursting (for instance, see the Hindmarsh-Rose oscillator \cite{cao2021}, the modified FitzHugh-Nagumo model considered in Ref. \cite{hua2022}, the FitzHugh-Rinzel model \cite{cao2021}, the pacemaker bursting neuron model analyzed in paper \cite{guan2020}, etc.) is considered as a most interesting issue.

\begin{acknowledgements}
The authors acknowledge support by the Russian Science Foundation (project No.  23-72-10040). 
\end{acknowledgements}

\section*{Declarations} 

\section*{Conflict of interest} The authors declare that they have no conflict of interest.

\section*{Data availability} The data that support the findings of this study are available from the corresponding author upon reasonable request.



\end{document}